\documentclass{llncs}

\usepackage{cite}

\usepackage{listings}
\usepackage[table]{xcolor}
\usepackage{xspace}
\usepackage{cite}
\usepackage[hyphens]{url}
\usepackage{flushend}
\usepackage{pgfplots, pgfplotstable}
\usetikzlibrary{patterns}
\usepackage{hyperref}
\usepackage{tikz}
\usepackage{paralist}
\usepackage{todonotes}
\usepackage{footnote}
\usepackage{framed}
\usepackage[]{algorithm2e}
\usepackage{adjustbox}
\usepackage{array}
\usepackage{multirow}

\pgfplotsset{compat=newest}

\newcommand{\tool}{GaLity}

\begin{document}

\title{Analyzing the Gadgets}
\subtitle{Towards a Metric to Measure Gadget Quality}

\author{Andreas Follner\inst{1}, Alexandre Bartel\inst{1} \and Eric Bodden\inst{2,3}\thanks{At the time this research was conducted Eric Bodden was at Fraunhofer SIT and TU Darmstadt.}}
\institute{Technische Universit\"at Darmstadt, Darmstadt, Germany \\ \email{andreas.follner@cased.de}, \email{alexandre.bartel@cased.de}
\and Paderborn University, Paderborn, Germany
\and Fraunhofer IEM, Paderborn, Germany  \\ \email{bodden@acm.org}}

\maketitle
\begin{abstract}
Current low-level exploits often rely on code-reuse, whereby short sections of code (\emph{gadgets}) are chained together into a coherent exploit that can be executed without the need to inject any code. Several protection mechanisms attempt to eliminate this attack vector by applying code transformations to reduce the number of available gadgets. Nevertheless, it has emerged that the residual gadgets can still be sufficient to conduct a successful attack. Crucially, the lack of a common metric for ``gadget quality'' hinders the effective comparison of current mitigations. 

This work proposes four metrics that assign scores to a set of gadgets, measuring quality, usefulness, and practicality. We apply these metrics to binaries produced when compiling programs for architectures implementing Intel's recent MPX CPU extensions. Our results demonstrate a 17\% increase in useful gadgets in MPX binaries, and a decrease in side-effects and preconditions, making them better suited for ROP attacks.

\end{abstract}

\section{Introduction}
\label{Sec:Intro}
Several mitigation techniques guarding against control-flow attacks have been developed over the past 15 years. In contrast to modern-day attacks~\cite{stitching_gadgets, coop, jit_code_reuse, reader_666, losing_control, cfi_bending, rop_dangerous, out_of_control, heap_feng_shui, antirop_evaluation}, the attacks of the 90s~\cite{aleph} were simple. The latter typically exploited a stack-based buffer overflow vulnerability to overwrite a stack frame's return address with another that points to a location at which the attacker had previously injected malicious code. On returning from the compromised function, execution would consequently be redirected to the injected code block.

Since the early 2000s, the prevalent processor architectures have adopted the \emph{No-eXecute} (NX bit) extensions. These allow an operating system to mark memory pages that only contain data (namely the heap and stack) as being non-executable~\cite{microsoft_dep}, thus stopping code-injection attacks.
However, programs may need to be able to allocate executable memory, for example for just-in-time compilation~\cite{aycock2003brief}. For such cases, the operating system provides several API calls that can change the memory protection level of a memory area (e.g., \texttt{VirtualProtect}\footnote{\url{https://msdn.microsoft.com/en-us/library/windows/desktop/aa366898\%28v=vs.85\%29.aspx}} on Windows).
These API calls were quickly abused by attackers, who would leverage them to change the access privileges of a region of memory where they had previously injected their payload.  
To circumvent the NX bit protection and to execute the API calls which change the memory protection of the payload code to executable, current exploits reuse executable code snippets, or \emph{gadgets},
comprising code from the running program and loaded libraries.
Such attacks are known as code-reuse attacks, the most popular and widespread technique being Return-Oriented Programming (ROP)~\cite{rop_org, rop}. 

The difficulty of staging a ROP attack in practice is subject to an attacker's concrete aims, the underlying environment, and the available gadgets. The latter, in particular, varies enormously between binaries. However, there is currently no established metric for quantifying the utility of gadgets within a given binary. Having such a metric would enable the comparison of
gadgets in various kinds of transformed binaries, e.g., 
different optimization levels of compilers, or binaries that have been rewritten to add instructions for exploit mitigation. Currently, many tools that produce such binaries, even those meant to enhance a binary's security, do not take into account how their transformation affects ROP gadgets. Especially for exploit-mitigation techniques this is counterproductive: if a mitigation technique transforms code, how does one know that it does not in the end increase a binary's attack surface by adding useful gadgets?

This work presents four metrics based on practical exploit development, that are designed to aid researchers in the evaluation of mitigations.
More generally, these metrics allow one to determine whether a binary transformation introduces gadgets that are better suited for ROP attacks than the original binary. 

Since it is somewhat difficult to make statements about the usefulness of a set of gadgets without knowing the goal of the attacker and the underlying environment, the metrics cover two targeted, real-world exploitation scenarios, and two more general computations which reflect gadget variety and gadget usability. This work further applies the metrics to binaries protected by MPX (Memory Protection eXtensions)~\cite{mpx}, a new buffer-overflow mitigation technique from Intel that adds instrumentation code to binaries through the compiler. As our evaluation shows, MPX-enabled binaries actually do contain more useful gadgets, and thereby increase the attack surface. This is particularly worrysome when running MPX-enabled binaries on legacy hardware that cannot benefit from the increased security that MPX is designed to offer. 
To summarize, our key contributions are:

\begin{itemize}
\item a definition of four metrics to measure gadget quality,
\item \tool{}, an open-source implementation to compute metrics on sets of gadgets, and
\item a case study using the metrics to determine how MPX affects gadgets on eight representative Windows x64 binaries.
\end{itemize}

The remainder of this paper is organized as follows. Section~\ref{Sec:Background} motivates the necessity to evaluate gadget quality. Section~\ref{Sec:EvalCrit} describes the proposed metrics. Section~\ref{Sec:Eval} explains the conducted case study on MPX. Section~\ref{Sec:Related} covers related work. Finally, Section~\ref{Sec:Conclusion} concludes the paper.

\section{Motivation}
\label{Sec:Background}
To the best of our knowledge, there exists no metric to assess the \emph{quality} of a gadget or a set of gadgets. Such a metric, however, has a large variety of use cases. For example, it could be used to compare different control-flow integrity (CFI)~\cite{cfi} approaches. Today's CFI implementations \cite{kbouncer, ropecker, cfi_for_cots, lockdown, forward_edge_cfi, ccfi} often use a metric which measures the reduction of gadgets (such as AIR, the average indirect target reduction~\cite{cfi_for_cots}, or DAIR, the dynamic average indirect target reduction~{\cite{lockdown}) to compare their results. For many approaches, this metric shows a reduction of over 99\%, yet this does not take into consideration the total number of gadgets, nor the quality of the remaining gadgets, limiting the metric's practical use. A DAIR of 90\% that leaves 50 gadgets with many side effects and preconditions intact is likely more secure than a DAIR of 99.5\%, that leaves intact exactly those 7 gadgets that an attacker requires to craft an exploit. Researchers using those metrics frequently acknowledge their limitations and  the difficulty of developing a metric that measures gadget quality~{\cite{cfi_bending, lockdown, forward_edge_cfi}}.

In general, attackers favour simple gadgets which have a minimum of side effects and preconditions.
For example, consider a gadget that loads the value that \texttt{rsp} points to into \texttt{rax}. A clean and effective gadget for achieving this would be: \texttt{pop rax ; ret}. In contrast, the gadget: \texttt{pop rax ; push rsp ; pop rbp ; mov [rdi+0x34fa], rsp ; ret 0x2dbf1} will also achieve this aim, but will also have the side-effect of overwriting \texttt{rbp}. In addition, this gadget has the precondition that \texttt{rdi+0x34fa} has to point to writeable memory. Finally, \texttt{ret 0x2dbf1} not only adds a large offset to \texttt{rsp} (which can be an issue if attacker-controlled memory is scarce, because it might set \texttt{rsp} to point outside of the allocated memory), it also disaligns the stack pointer, which is something normal programs do not do, hinting at a possible exploit execution.
The next Section presents the four metrics we propose to compute gadget quality.

\section{Metrics for Measuring Gadget Quality}
\label{Sec:EvalCrit}
In general, evaluating the quality of a set of gadgets is non-trivial. This stems primarily from the fact that an attacker's goal is potentially unknown, and that given sufficient gadgets, one can construct practically any program. In addition, the gadgets required for an attacker to achieve a goal vary by operating system and architecture. For example, on Windows x86, parameters to functions are usually passed on the stack, while on Windows x64, the first four parameters are passed through registers and all remaining ones are passed on the stack\footnote{\url{https://msdn.microsoft.com/en-us/library/windows/hardware/ff561499\%28v=vs.85\%29.aspx}}, leading to differences in gadget requirements.
As a running example, we consider exploits targeting \texttt{VirtualProtect}, which is an API call that commonly serves as an avenue to bypassing NX protection on Windows 7 x64~\cite{CVE-2015-0349, CVE-2013-3893, blend_exploit}. We stress that our four metrics are not bound to evaluating this specific API call, as they consider the more general attack setup and execution procedures associated with ROP exploits. In addition, we perform an in-depth analysis of the various properties of gadgets with respect to their side effects, preconditions, usability, and usefulness.

\subsection{Metric 1: Gadget Distribution}
\label{sec:gadget_distribution}

The \emph{gadget distribution} metric is calculated by partitioning a given set of gadgets into twelve broad categories, with each category representing a class of operations, such as \emph{arithmetic} and \emph{data move}, as shown in Table~\ref{tab:gadget_cats}. Gadgets are assigned to a category based on the first instruction of a gadget. 
For example, the gadget \texttt{add rax, 0x40 ; pop rcx ; ret} would be assigned to the \emph{arithmetic} category. We categorize on the basis of the first instruction as every suffix of a gadget is itself a gadget, and will be categorized separately.
Note that gadgets containing privileged or sensitive instructions~\cite{intel_architecture_manual} are discarded and not considered in further steps because they trap in user mode, thereby making a gadget unusable.

Analyzing the distribution of frequencies of gadgets amongst categories is helpful as it allows comparing whether the distribution of gadgets in a transformed binary is similar to the one in the original binary, or if the number of gadgets in a category useful for an attacker has grown. Gadget quality and usefulness, however, are not measured and addressed by the remaining metrics.

While Table~\ref{tab:gadget_cats} does not contain all instructions of the x86-64 instruction set, it covers 99\% of the instructions found in gadgets of the binaries we used in the evaluation, i.e., a total of 20 MiB containing over one million instructions.
Due to the large size of the x86-64 instruction set (over 700 instructions~\cite{intel_architecture_manual}), it would be a time-consuming, manual process to cover all existing instructions. However, the fact that we do not achieve 100\% coverage does not pose a threat to the metric, because all important and common instructions are categorized. The few we we did not include do not have a big impact on the overall distribution. A manual inspection of uncategorized instructions in other binaries (we used several Windows 7 system libraries) revealed that there were many different instructions but in small numbers in any of the inspected binaries, which is what we expected.

~\textit{Metric 1 allows to assess whether a transformed binary contains more gadgets in categories useful to an attacker.}

\newcommand{\rdst}[0]{\ensuremath{r_d}\xspace}

\begin{center}
\begin{savenotes}
\begin{table}
    \caption{Gadget Categories}
    \begin{tabular}{ | l | p{9.75cm} |}
    \hline
    \bf Category & \bf Included Instructions \\ \hline
    Data move & \texttt{pop, push, mov, xchg, lea, cmov, movabs} \\ \hline
    Arithmetic & \texttt{add, sub, inc, dec, sbb, adc, mul, div, imul, idiv, xor, neg, not\footnote{It might appear peculiar that \texttt{xor, neg, not} are in the arithmetic category - however, this is how exploit developers often use these instructions. Since using nullbytes is sometimes prohibited by the environment, writing the negated or xor-ed value in memory, loading it to a register and then using the same operation on it again is used to bypass this restriction.}} \\ \hline
    Logic & \texttt{cmp, and, or, test} \\ \hline
    Control flow & \texttt{call, sysenter, enter, int, jmp, je, jne, jo, jp, js, lcall, ljmp, jg, jge, ja, jae, jb, jbe, jl, jle, jno, jnp, jns, loop, jrcxz} \\ \hline
    Shift \& Rotate & \texttt{shl, shr, sar, sal, ror, rol, rcr, rcl} \\ \hline
    Setting flags & \texttt{xlatb, std, stc, lahf, cwde, cmc, cld, clc, cdq} \\ \hline
    String & \texttt{stosd, stosb, scas, salc, sahf, lods, movs} \\ \hline
    Floating point & \texttt{divps, mulps, movups, movaps, addps, rcpss, sqrtss, maxps, minps, andps, orps, xorps, cmpps, vsubpd, vpsubsb, vmulss, vminsd, ucomiss, subss, subps, subsd, divss, addss, addsd, cvtpi2ps, cvtps2pd, cvtsd2ss, cvtsi2sd, cvtsi2ss, cvtss2sd, mulsd, mulss, fmul, fdiv, fcomp, fadd}  \\ \hline
    Misc & \texttt{wait, set, leave}  \\ \hline
    MMX & \texttt{pxor, movd, movq} \\ \hline
    NOP & \texttt{nop} \\ \hline
    RET & \texttt{ret} \\ \hline
    \end{tabular}
    \label{tab:gadget_cats}
\end{table}
\end{savenotes}
\end{center}

\subsection{Metric 2: Gadget Environment Setup Capabilities}

When constructing a ROP exploit, an attacker must be able to prepare the environment and operands for subsequent gadgets in a chain. For example, when attempting to perform a Windows API call via ROP, an attacker will generally require the ability to specify the call's arguments. The degree of ease with which an attacker may manipulate memory will affect the choice of gadgets that she uses. In this metric, we consider the most general case, whereby an attacker is able to inject arbitrary arguments into a target program's memory space at a known location. This could be possible due to, e.g., a browser with Javascript turned on, allowing heap sprays and Heap Feng Shui~\cite{heap_feng_shui}, and other vulnerabilities like information leaks~\cite{info_leakage}. We further assume the vulnerable program is running on a Windows~7 x64 machine, which is a very common platform.

Consider the case whereby an attacker wants to invoke \texttt{VirtualProtect}, which takes four arguments. On the aforementioned target platform, the first four parameters are passed through registers (\texttt{rcx, rdx, r8, r9}). In such a scenario, an attacker needs to make sure that those registers contain the correct values before \texttt{VirtualProtect} can be invoked. To achieve that, three different kinds of gadgets are required, namely:
\begin{inparaenum}[(i)]
\item a \emph{stack pivot} gadget which points \texttt{rsp} to the injected data, i.e., function arguments and addresses of gadgets,
\item gadgets to load the arguments from memory to the appropriate registers, and
\item a gadget that calls \texttt{VirtualProtect}.
\end{inparaenum} 

This metric looks for gadgets that achieve these goals and distinguishes between gadgets that achieve only the required task or include other instructions. 
Of course, our tool reports gadgets only if the register that receives the argument is preserved, i.e., not overwritten by another instruction in the same gadget. In case the attacker wants to invoke an API that requires fewer arguments, like \texttt{VirtualAlloc}\footnote{\url{https://msdn.microsoft.com/en-us/library/windows/desktop/aa366887\%28v=vs.85\%29.aspx}}, fewer gadgets that load arguments are required.
\newline
A gadget is only useful in preparing a destination register \rdst for use within a ROP chain if it does not destroy its value prior to returning. More concretely, consider a gadget consisting of a sequence of $n$ instructions $i_0 \mathtt{;\,} i_1 \mathtt{;\,} \dots \, i_{n - 1} \mathtt{;\,} \mathtt{ret}$. If $i_0$ assigns the value to \rdst, any subsequent instruction $i_k$ with $k > 0$ that has \rdst as a target operand and falls within the \emph{data move}, \emph{arithmetic}, or \emph{shift and rotate} categories is tagged as being potentially destructive. A second refinement step is subsequently carried out, whereby the quirks of the target architecture are taken into account. For instance, instructions that output to a 32 bit subregister are handled differently than those that output to 16 or 8 bit subregisters. This is due to the behaviour that writing to a 32 bit subregister automatically zero-extends the value to fill the entire 64 bit register~\cite{intel_architecture_manual}.

In the case of exploits making use of \texttt{VirtualProtect}, one finds that three of the four arguments that this API call takes (namely \texttt{lpAddress}, the start address of the memory region whose protection level is to be changed, \texttt{dsSize}, the size address of the memory region whose protection level is to be changed, and \texttt{lpflOldProtect}, an address where the old protection level will be stored) do not need to be precise. If \texttt{lpAddress} is a few bytes off an attacker can take this into account, just like a slightly smaller or larger size argument. \texttt{lpflOldProtect} is not used by an attacker and can therefore be written to any location. Therefore, the metric only deems two instructions destructive, namely \texttt{pop} and \texttt{mov} in 64 bit or 32 bit subregisters, as they overwrite the whole register.
~\textit{Metric 2 allows one to assess whether a transformed binary contains gadgets typically required for an attack where the environment gives the attacker a lot of leeway.}

\subsection{Metric 3: Gadget Environment Setup Capabilities - Restricted}
In contrast to the previous metric, this metric considers the case where an attacker is restricted in the ways in which she can inject values into memory. In particular, we consider the scenario where an attacker may only inject data and hijack the control-flow via \texttt{strcpy}. This complicates the direct injection of values into memory because many parameters to API calls often contain null-bytes, which terminate strings, thus requiring that the arguments to be used for correctly invoking a function such as \texttt{VirtualProtect} be calculated dynamically at runtime.
Imagine an attacker wants to indeed invoke \texttt{VirtualProtect}. By taking a look at the required parameters it becomes clear that many will contain null-bytes: \texttt{lpAddress} should point to the payload. Depending on the memory layout, this address may contain null-bytes (e.g., in a classic stack buffer overflow vulnerability on Windows, stacks are located at very low addresses making it very likely for the address to have its leftmost bytes set to null). 
\texttt{dwSize} must not be too large, i.e., \texttt{lpAddress + dwSize} must include only mapped pages. The value must also not be too small, as it has to cover the memory area where the payload is injected. Typically, the value is a couple of thousand bytes or smaller, which is a value that cannot be injected directly. \texttt{flNewProtect} is usually set to \texttt{0x40}, which cannot be injected directly because the leftmost bytes are null,and requires to be computed at runtime. 
\texttt{lpflOldProtect} will receive the old protection value, hence must point to writable memory, which may contain null-bytes. 
This example shows that in a scenario where the attacker is restricted, she will require various arithmetic and data-move gadgets in order to dynamically calculate parameters for API calls using gadgets.

The metric gauges the presence of gadgets that may be used to assist in evaluating values dynamically at runtime, specifically gadgets that move data between memory and registers and compute values: \texttt{pop}, \texttt{push}, \texttt{add}, \texttt{sub}, \texttt{adc}, \texttt{dec}, \texttt{inc}, \texttt{neg}, \texttt{not}, \texttt{mov}, \texttt{sbb}, \texttt{xchg}, \texttt{xor}. As in the case of \textit{Metric 2}, a gadget is only considered if \rdst is preserved.
~\textit{Metric 3 allows one to assess whether a transformed binary contains gadgets typically required for an attack where the attacker has to make many calculations at runtime and cannot inject arbitrary data into a program.}

\subsection{Metric 4: Gadget Quality}
\label{sec:gadqual}
The aforementioned metrics do not measure the quality of a gadget per se, rather they provide an indication whether a specific attack can succeed given a set of gadgets. This metric focuses on assessing the quality of an individual gadget, whereby a high-quality gadget is defined as one having no preconditions or side-effects on other registers or memory. An example of a precondition is that a specific register has to point to writeable memory,
e.g., in the gadget \texttt{pop rax ; mov [rdi+0x34fa], rsp ; ret}. To be usable, \texttt{rdi+0x34fa} must point to writeable memory. A side-effect is, for example, that data in another register is overwritten or the stack pointer is manipulated in a way that is difficult to undo, e.g., in the gadget \texttt{pop rax ; mov rcx, 0xb0adffff ; leave ; ret}. This gadget overwrites the values in \texttt{rcx}, \texttt{rsp}, and \texttt{rbp}. To express gadget quality, a score is calculated for every gadget considered useful (see \textit{Metric~3}). The score starts at 0 and is increased for side-effects and preconditions. Therefore, a higher score equals worse gadget quality. In the following we give a high-level overview of the two criteria we use to calculate the score for gadget quality.

\subsubsection*{Grading Instructions}
To measure side-effects and preconditions, the metric inspects every instruction in a gadget. It reuses the categories introduced in Section~\ref{sec:gadget_distribution} and assigns a score to each category, which reflects how destructive the instructions in the respective category are. 
Table~\ref{tab:score} summarizes the scoring system.
Depending on the destination of the instruction, we apply a modifier to the originally assigned score. The metric recognizes three possible kinds of destinations:~\texttt{rsp}, which should ideally not be modified, because it is responsible for the control flow and always needs to point to the next gadget. Therefore, modifications of \texttt{rsp} usually have the largest influence on the overall score of a gadget. The second possible destination is $r_d$, the destination register in the first instruction of a gadget, for which we assume that this is also the register an exploit developer is interested in not being modified later on in the same gadget (in case a memory address is the target there is no active register; in case of an \texttt{xchg} instruction, both registers are active registers). Modifications of $r_d$ are generally not desirable, but, depending on the modification, can be reversible, e.g., simple arithmetic. The third possible destination is any other general purpose register, except \texttt{rsp} and $r_d$, the metric considers all undesirable side effects and preconditions. Even if they do not affect \texttt{rsp} or $r_d$ directly, they still negatively impact the final score.

\begin{table}[h!]
    \caption{Rules for grading instructions. Category describes the category of the instruction (see Table~\ref{tab:gadget_cats}). ``RSP'', ``rd'' and ``Other'' are possible targets for instructions, the stack pointer, the destination register of the first instruction of a gadget, or any of the other general purpose registers respectively. Categories not in the table generally do not affect the score, with some exceptions discussed in Section~\ref{sec:gadqual}}
    \begin{tabular}{ | c | c | c | c | p{10cm} | }
    \hline
     \rotatebox[origin=c]{90}{\bf Cat.} & 
     \rotatebox[origin=c]{90}{\bf RSP} & 
     \rotatebox[origin=c]{90}{\bf rd} & 
     \rotatebox[origin=c]{90}{\bf Other} & 
     \multicolumn{1}{>{\centering\arraybackslash}m{10cm}|}{\bf Notes} \\ \hline
    \multirow{1}{*}{\rotatebox[origin=c]{90}{\parbox{1.1cm}{Data\\move}}} & \multicolumn{1}{>{\centering\arraybackslash}m{.25cm}|}{2} & \multicolumn{1}{>{\centering\arraybackslash}m{.25cm}|}{1} & \multicolumn{1}{>{\centering\arraybackslash}m{.25cm}|}{0.5} & As opposed to all other instructions in this category, \texttt{push} does not affect the score of a gadget, since the only side effect it has is on \texttt{rsp}, and changes to \texttt{rsp} are covered by our \texttt{rsp} monitoring. \\ \hline
    \multirow{1}{*}{\rotatebox[origin=l]{90}{\parbox{2cm}{Arithmetic}}} & 2 & 1 & 0.5 & Arithmetic instructions that modify a register other than \texttt{rsp} can be taken into account by the exploit developer. E.g., if \texttt{r8} should contain \texttt{0x40}, and a gadget like \texttt{pop r9 ; add r8, 0x10 ; ret} has to be executed as the last gadget, the developer can simply make sure \texttt{r8} contains the value \texttt{0x30} before invoking the last gadget. Arithmetic instructions modifying \texttt{rsp} are covered by our \texttt{rsp} monitoring. \\ \hline
    \multirow{1}{*}{\rotatebox[origin=c]{90}{\parbox{1.1cm}{Shift \&\\ Rotate}}} & 3 & 2 & 0.5 & These instructions are handled similarly to arithmetic instructions, however, they are more difficult to take into account, which is why they increase the score more than arithmetic instructions. \\ \hline
    \end{tabular}
    \label{tab:score} 
\end{table}

In a few cases grading all instructions in one category the same does not make sense and would result in false scoring, which is the reason for the following exceptions. \textit{Exception \#1:} Certain instructions that modify \texttt{rsp} need to be treated differently. This covers all instructions where we can statically determine the offset applied to \texttt{rsp}. Depending on how much \texttt{rsp} is changed, we adjust the overall score of the gadget. The details on this are covered in the next subsection. In case it is not possible to statically determine the offset (e.g., \texttt{leave} or \texttt{pop rsp}), the overall score of the gadget is increased depending on the category of the instruction, as presented in Table~\ref{tab:score}. \textit{Exception \#2:} \texttt{leave} is the only instruction in the miscellaneous category that needs to be taken into account, as it affects \texttt{rsp}. This is taken care of through our \texttt{rsp} monitoring. \textit{Exception \#3:} 
Remember from Section~\ref{sec:gadget_distribution} that we do not cover all of the x86-64 instructions. 
This means that in very rare cases (less than 0.1\%) we cannot grade a gadget because it contains an instruction which we did not categorize. We discard these gadgets from the analysis. \textit{Exception \#4:} If an instruction uses a dereferenced register as destination its score is increased according to the rules in Table~\ref{tab:score}, because this poses a precondition - e.g., the gadget \texttt{pop r8 ; mov [rdx], 0xfffa ; ret} has the precondition that \texttt{rdx} has to point to writable memory before the gadget can be used.

\subsubsection*{Monitoring \texttt{rsp} Offset}
Modifications to \texttt{rsp} need to be tracked for each gadget, as explained in the previous paragraph. A short example will make clear why this is necessary. Assume the following gadget: \texttt{pop rax ; add rbx, 0x10ff ; push rcx ; ret}. In this case, \texttt{rsp} will point to the value  contained in \texttt{rcx} and jump to this address, which is not the injected address of the next gadget.
For keeping track of the \texttt{rsp} offset the metric uses an SP-Score, \textit{SPS}, which starts at 0, is increased for \texttt{pop} and decreased for \texttt{push} and \texttt{ret \textit{n}} instructions. Of course, also arithmetic instructions on \texttt{rsp} are monitored and the respective value is added to or subtracted from \textit{SPS}. When all instructions in a gadget have been analyzed and \textit{SPS} is not 0 this means that \texttt{rsp} does not point to the next gadget, which might be problematic. Therefore, if \textit{SPS} is negative, the overall score of the gadget will be increased by 2. Also, if \textit{SPS} is large (more than 4 KiB) or not aligned, the score of the gadget will be increased by 1, because the former requires an attacker to be able to control more memory and the latter can be detected easily by exploit mitigation tools. If the instruction that operates on \texttt{rsp} takes a register and not an immediate (e.g., a \texttt{add rsp, rcx}), \textit{SPS} is not changed but the gadget score will be increased by rules in Table~\ref{tab:score}.
~\textit{Metric 4 allows one to assess the overall ``quality'' of a set of gadgets in respect to side-effects, preconditions, and usability.}

\subsection{Discussion of the Metrics}
We believe that metrics that measure the quality of a set of gadgets should focus on practical relevance rather than a theoretical concept such as Turing completeness~\cite{rop_org}. Furthermore, they should also reflect whether real-world exploits can be constructed. Since at least Microsoft has seen a shift from classic, stack-based vulnerabilities to heap-related vulnerabilities~\cite{msir16}, we believe that metrics should still consider both of these classes of attacks. Last but not least, the metrics should not be limited to well-defined and realistic attack scenarios, but also express overall gadget quality, i.e., side-effects and preconditions. To summarize, metrics as described above should:
\begin{itemize}
\item Be practical
\item Measure if popular current attacks are possible with a given set of gadgets
\item Measure if popular past attacks are possible with a given set of gadgets
\item Measure gadget ``quality''
\end{itemize}
The proposed metrics achieve all these goals.
We would like to stress that our aim is assessing whether a binary contains gadgets suitable for today's ROP attacks. 
Recently, attacks that use longer and more complex gadgets have been proposed by researchers~\cite{rop_dangerous, stitching_gadgets, size_matters, out_of_control, antirop_evaluation}. 
Such attacks are designed to bypass specific mitigation techniques, which are not used in the real world. Thus, in current environments, these complex attacks are cumbersome as they offer no advantage over using regular and simpler ROP gadgets, and we are not aware of any of these complex attacks being used in the wild.}

Because of the lack of practical relevance, we decided not to treat gadgets potentially useful in such complex attacks differently than the other gadgets. 
Nevertheless, if new mitigations limiting the gadgets an attacker may use become widespread and attackers are forced to use more complex and longer gadgets and start using tools that assist in finding gadgets semantically rather than through simple pattern matching, our metrics will have to be updated to reflect this new environment.
This is why we also plan to use a more abstract interpretation of gadgets and look into leveraging synergies created by combining gadgets in the future. 
Furthermore, we also leave an extension to jump-oriented programming (JOP)~\cite{jop, jop2} for future work.

\section{Evaluation}
\label{Sec:Eval}
We have implemented the described metrics in a tool named \tool{}, which takes a textfile containing gadgets as input and outputs the metrics we described in Section~\ref{Sec:EvalCrit}. We demonstrate that it is both practical and useful by applying it to binaries that are compiled to use MPX~\cite{mpx}, Intel's latest mitigation technique against runtime exploits. MPX introduces new registers that contain the lower and upper bound of a pointer, and instructions that operate on those registers. This enables compilers to emit additional instructions (MPX and non-MPX) that tracks the sizes of buffers and accesses to those buffers at runtime, which can prevent buffer overflows. On processors which do not support MPX, MPX instructions execute as \texttt{nop}, making MPX compatible with older CPUs, but leaving those binaries unprotected by MPX. Given this observation one thus must wonder if the increased code size and thus increased availability of gadgets might actually decrease a binary's security on such systems. We then compare the results obtained by applying \tool{} to binaries compiled with MPX support with the results obtained by applying \tool{} to the same binaries compiled without MPX support, and determine which binaries contain more helpful gadgets for an attacker according to our metrics.

\subsection{Implementation}
We wrote \tool{} in C\#.
\tool{} takes a simple text file that contains gadgets as input and parses it in four passes, which correlate to the four metrics described in the previous section.  While doing everything in one pass is certainly possible, we decided to use several passes, as this increases code readability, and performance was no issue (even large sets of gadgets containing hundreds of thousands of gadgets can be analyzed in less than 10 seconds on an Intel Core 2 Duo with 4 GiB RAM). 
Since current ROP attacks use rather simple gadgets
we only reconstruct the semantic we require for our metrics. 
For example, \tool{} recognizes the differences between instructions outputting to 64 bit, 32 bit, 16 bit, or 8 bit (sub)registers and treats them accordingly, but does not recognize that many instructions manipulate CPU flags. Knowledge about this would be required when utilizing more complex gadgets that use conditional branches.
However, for current real-world attacks, the simpler but less error-prone approach is sufficient.

We looked into using an intermediate representation (IR) which makes side effects explicit, as this would allow more precise grading. However, we discovered that, as today's attacks use simple gadgets, there are few side effects that are relevant in our scenarios. Therefore, we leave designing an IR tailored to the very specific requirements of measuring gadget quality, that (1) can be reused and (2) recognizes more side effects, for future work.

\subsection{Setup}
To discover gadgets and write them to a file we used ROPgadget 5.4\footnote{\url{https://github.com/JonathanSalwan/ROPgadget}}, with a maximum gadget length of 15. This might sound like a very high number, however, we did not want to risk potentially missing some useful gadgets. Also, our metrics ensure that gadgets that do not preserve $r_d$ are discarded, i.e., not considered in the results, and gadgets that have many side effects have a bad score. Also, for this specific case study we decided to consider duplicate gadgets and not just unique gadgets, because if an important gadget exists several times in a binary, this binary is more attractive to an attacker than a binary which contains only one copy of that gadget. This matters, for example, in a scenario where a patch (security-related or not) or any other program modification removes said gadget. Furthermore, taking duplicate gadgets into account helps us measure, if the additional gadgets introduced by MPX are copies of useless or useful gadgets.

We compiled programs taken from SPEC2006, using Intel's latest GCC release with MPX support at the time of writing (5.0.0).\footnote{\url{https://software.intel.com/en-us/articles/intel-software-development-emulator}} 
We decided to use the SPEC suite because it covers a wide range of application types, and present parts of real programs. MPX is still new and not integrated too well in build chains, which made compiling any program a challenge. However, we got the following eight programs to work properly: 401.bzip2, 403.gcc, 435.gromacs, 456.hmmer, 458.sjeng, 464.h264ref, 473.astar, 482.sphinx3. We compiled all binaries four times, with and without MPX and with and without optimizations (-O2). However, for our evaluation we only considered optimized binaries as this reflects real-world binaries.

\subsection{Results}

\begin{table}[ht]
    \caption{Results for \textit{Metrics~2},~\textit{3}, and~\textit{4}. Columns rcx, rdx, r8 and r9 denote the number of gadgets which load a value in the respective register, column pivot denotes the number of stack pivot gadgets. The first number denotes the number of gadgets without side-effects, the second number the number of gadgets with side-effects. Column call denotes the number of gadgets usable for indirect calls. These numbers are required for computing \textit{Metric~2}. Column useful denotes the number of useful gadgets, calculated by \textit{Metric~3}. Column Q denotes the number of gadgets with a score of 1 or lower, calculated by \textit{Metric~4}.}
   \resizebox{\textwidth}{!}{
    \begin{tabular}{ | l | p{1.3cm} | p{1.3cm} | p{1cm} | p{1cm} | p{1.2cm} | p{0.8cm} | p{1.25cm} | p{1.25cm} |}
    \cline{2-9}
	\multicolumn{1}{c|}{ } & \multicolumn{6}{c|}{\textit{Metric 2}} & \multicolumn{1}{c|}{\textit{Metric 3}} & \multicolumn{1}{c|}{\textit{Metric 4}} \\ \hline
    \bf Program  & \bf rcx & \bf rdx & \bf r8 & \bf r9 & \bf pivot & \bf call & \bf useful & \bf Q \\ \hline
    \rowcolor{lightgray}
    h264ref no MPX & 4 / 29 & 1 / 8  & 1 / 9 & 0 / 0 & 0 / 453 & 62 & 6,056 & 3,749 \\ \hline
    \rowcolor{lightgray}
    h264ref MPX & 7 / 29 & 0 / 23  & 1 / 3 & 0 / 1 & 0 / 666 & 91 & 7,546 & 4,906 \\ \hline
    gromacs no MPX & 228 / 320 & 39 / 135  & 0 / 2 & 0 / 0 & 0 / 1071 & 84 & 10,823 & 6,563 \\ \hline
    gromacs MPX & 228 / 418 & 36 / 141 & 0 / 7 & 0 / 1 & 0 / 1214 & 155 & 13,002 & 8,170 \\ \hline
    \rowcolor{lightgray}
    hmmer no MPX & 6 / 24 & 3 / 27 & 0 / 3 & 0 / 0 & 0 / 509 & 33 & 5,539 & 3,303 \\ \hline
    \rowcolor{lightgray}
    hmmer MPX & 8 / 21 & 4 / 19 & 0 / 2 & 0 / 0 & 0 / 469 & 39 & 6,188 & 3,952 \\ \hline
    gcc no MPX & 4 / 71 & 2 / 219 & 0 / 14 & 0 / 8 & 6 / 5295 & 588 & 50,766 & 32,949 \\ \hline
    gcc MPX & 2 / 52 & 4 / 71 & 0 / 9 & 0 / 4 & 0 / 4337 & 763 & 59,522 & 39,342 \\ \hline
    \rowcolor{lightgray}
    sphinx3 no MPX & 2 / 14 & 0 / 11 & 0 / 0 & 0 / 0 & 0 / 230 & 29 & 3,189 & 1,964 \\ \hline
    \rowcolor{lightgray}
    sphinx3 MPX & 1 / 11 & 0 / 7 & 0 / 0 & 0 / 0 & 1 / 251 & 52 & 3,484 & 2,323 \\ \hline
    sjeng no MPX & 1 / 3 & 0 / 3 & 0 / 0 & 0 / 1 & 0 / 122 & 72 & 1,444 & 983 \\ \hline
    sjeng MPX & 1 / 4 & 0 / 5 & 0 / 0 & 0 / 0 & 0 / 137 & 76 & 1,982 & 1,414 \\ \hline
    \rowcolor{lightgray}
    astar no MPX & 1 / 4 & 0 / 4 & 0 / 0 & 0 / 0 & 0 / 122 & 11 & 1,009 & 584 \\ \hline
    \rowcolor{lightgray}
    astar MPX & 0 / 5 & 0 / 2 & 0 / 0 & 0 / 0 & 0 / 140 & 12 & 1,203 & 698 \\ \hline
    bzip2 no MPX & 0 / 1 & 0 / 1 & 0 / 0 & 0 / 0 & 0 / 99 & 13 & 790 & 466 \\ \hline
    bzip2 MPX & 0 / 1 & 0 / 1 & 0 / 0 & 0 / 0 & 0 / 112 & 16 & 987 & 605 \\ \hline
    \end{tabular}
    }
    \label{tab:res} 
\end{table}

First of all, we noticed that MPX has a big influence on file size. With no optimizations, an MPX binary is, on average, almost 3 times as large as a non-MPX binary. With optimization level~2, which we used throughout our experiments, an MPX binary is still, on average, 86\% larger compared to a non-MPX binary. We noticed that, while the file size increases by a factor of almost two, the number of gadgets does not increase in the same way, MPX binaries contain, on average, only 23\% more gadgets than non-MPX binaries. This is because the number of gadgets is directly related to the number of \texttt{ret} instructions in a binary. MPX does not add many new functions but rather makes existing functions longer, therefore only few intended new \texttt{ret} instructions appear. Unintended \texttt{ret} instructions~\cite{rop} might appear in some cases, however, since the new opcodes introduced by MPX do not contain a \texttt{ret} opcode, the possibility for this is rather low.

\begin{figure}
\begin{tikzpicture}
\begin{axis}[
    title = {},
    width=12.25cm,
    height=7cm,
    xtick={1,...,11},
    xticklabels={%
        Arithmetic,
        Data-Move,
        Control-Flow,
        Logic,
        RETs,
        Shift/Rot,
        Flag,
        String,
        NOP,
        FP,
        Misc},
        x tick label style={rotate=45, anchor=north east, inner sep=0mm},
        ylabel={Relative gadget distribution in \%},
    grid=major,
    ybar
    ]

\addplot[
    fill=blue!25,
    draw=black,
    point meta=y,
    every node near coord/.style={inner ysep=5pt},
    error bars/.cd,
        y dir=both,
        y explicit
] 
table [y error=error] {
x   y           error    label
1   27.85   	2.23		1
2	31.67		4.73		2
3	5.18		0.85		3
4	6.63		1.37		4
5	20.60		2.34		5
6	2.70		0.30		6
7	1.19		0.44		7
8	0.78		0.32		8
9	1.10		0.33		9
10	1.28		0.65		10
11	0.27		0.11		11
};

\addplot[
    fill=red!25,
    draw=black,
    point meta=y,
    every node near coord/.style={inner ysep=5pt},
    error bars/.cd,
        y dir=both,
        y explicit
] 
table [y error=error] {
x   y           error    label
1	26.97		2.86		1
2	29.10		5.15		2
3	5.24		1.38		3
4	5.73		0.81		4
5	25.60		2.53		5
6	2.25		0.36		6
7	0.93		0.43		7
8	0.48		0.13		8
9	2.03		0.44		9
10	1.00		0.42		10
11	0.16		0.07		11
};

\draw ({rel axis cs:0,0}|-{axis cs:0,0}) -- ({rel axis cs:1,0}|-{axis cs:0,0});
\end{axis}
\end{tikzpicture}
\caption{This figure shows the average distribution of gadgets across all eight applications. The blue bar represents the non-MPX binaries, while the red bar represents the MPX binaries.}
\label{fig:dis}
\end{figure}
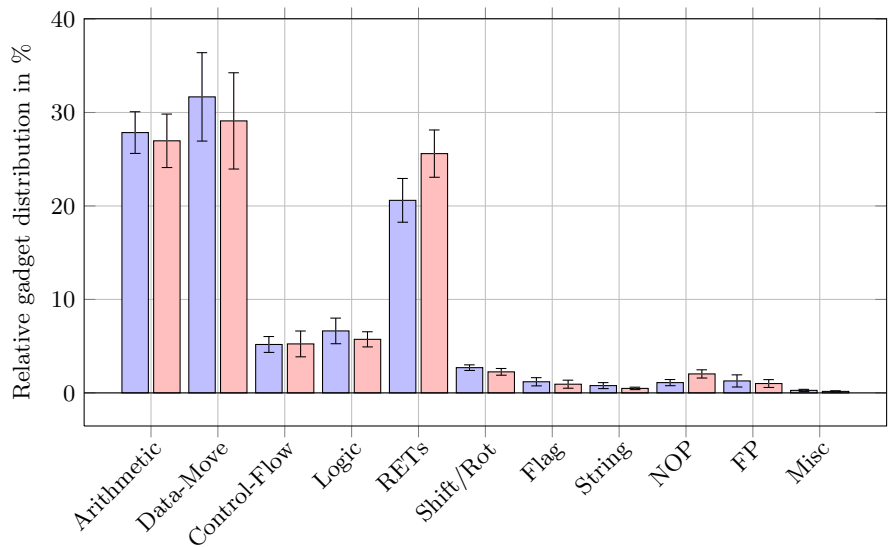

Analyzing the increase or decrease of gadgets for each category due to MPX, illustrated in Figure~\ref{fig:growth}, shows that most categories gain gadgets. Arithmetic gadgets, which are helpful to an attacker, increase in both number and diversity. Data-move gadgets grow in numbers, but do not change a lot in respect to diversity. An interesting observation is that NOP-gadgets increase drastically, which is presumably due to the fact that the new MPX instructions are interpreted as multi-byte \texttt{NOP}s on hardware that does not support MPX. The categories flag, string and floating-point have a high standard deviation, indicating that changes in these categories are very application-specific. Gadgets in the miscellaneous category decrease both in diversity and number. Despite the large increase of \texttt{nop} gadgets, the overall distribution of gadgets remains roughly the same, as Figure~\ref{fig:dis} shows. Overall we conclude that MPX binaries contain more gadgets in categories helpful to an attacker.

\begin{figure}
\begin{tikzpicture}
\begin{axis}[
    title = { },
    width=12.25cm,
    height=7cm,
    xtick={1,...,11},
    xticklabels={
        Arithmetic,
        Data-Move,
        Control-Flow,
        Logic,
        RETs,
        Shift/Rot,
        Flag,
        String,
        NOP,
        FP,
        Misc},
        x tick label style={rotate=45, anchor=north east, inner sep=0mm},
        ylabel={Increase / decrease for each category in \%},
    grid=major,
    ybar
    ]

\addplot[
    fill=blue!25,
    draw=black,
    point meta=y,
    every node near coord/.style={inner ysep=5pt},
    error bars/.cd,
        y dir=both,
        y explicit
] 
table [y error=error] {
x   y           error    label
1   14.01   	10.84		1
2	-5.35		15.02		2
3	24.07		28.13		3
4	2.76		22.07		4
5	-3.94		9.73		5
6	-0.36		17.22		6
7	13.98		75.05		7
8	33.44		139.17		8
9	147.38		48.51		9
10	-3.96		21.24		10
11	-17.16		11.50		11
};

\addplot[
    fill=red!25,
    draw=black,
    point meta=y,
    every node near coord/.style={inner ysep=5pt},
    error bars/.cd,
        y dir=both,
        y explicit
] 
table [y error=error] {
x   y           error    label
1	22.42		10.01		1
2	15.89		9.75		2
3	28.23		25.68		3
4	13.42		28.34		4
5	57.76		14.24		5
6	5.11		13.48		6
7	20.54		80.23		7
8	28.11		141.51		8
9	142.25		43.81		9
10	5.91		25.03		10
11	-25.34		16.64		11
};

\draw ({rel axis cs:0,0}|-{axis cs:0,0}) -- ({rel axis cs:1,0}|-{axis cs:0,0});
\end{axis}
\end{tikzpicture}
\caption{This figure shows the average growth of gadgets for each category due to MPX across all eight applications. The blue bar represents the increase considering only unique gadgets, while the red bar represents the total increase of gadgets, i.e., also duplicate gadgets. We use the information about how the number of unique gadgets changes to infer if and how gadget variety is affected by a program transformation.}
\label{fig:growth}
\end{figure}
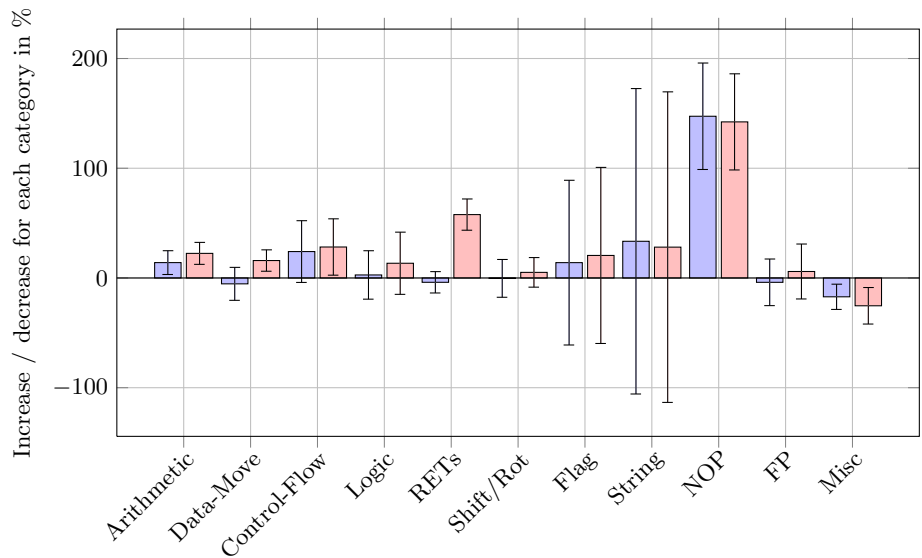

Next, we are interested in the two attack scenarios, i.e., \textit{Metrics~2} and~\textit{3}. Regarding \textit{Metric~2}, there is no big difference in the availability of gadgets. Gadgets that load arguments in \texttt{r8} or \texttt{r9} are rare in both MPX and non-MPX binaries, and sometimes the MPX binary and sometimes the non-MPX binary contains some. Regarding \textit{Metric~3}, the number of useful gadgets increases in every binary and on average by 17\%, making MPX binaries a much more attractive target to attackers. We summarize the results in Table~\ref{tab:res}.
Lastly, we determine overall gadget quality using \textit{Metric~4}. In all eight binaries, the MPX versions contain more gadgets of high quality, i.e., with fewer side-effects and preconditions, as the last column of Table~\ref{tab:res} shows.

By taking all four results into consideration we come to the conclusion, that binaries compiled with MPX support are favourable for an attacker. \textit{Metric~1} shows an overall increase of gadgets in useful categories, further confirmed by \textit{Metric~3}, which also shows that the additional gadgets in those categories are useful in practice. \textit{Metric~2} gives no indication that MPX or non-MPX binaries contain more of the required gadgets. \textit{Metric~4} gives the indication that MPX binaries tend to have more gadgets of higher quality, making them easier to use for an attacker.

\section{Related Work}
\label{Sec:Related}
To the best of our knowledge, no previous work has been done on the topic of designing a metric to measure the quality of a set of gadgets, even though the metrics currently used to measure CFI strength are insufficient, exactly because gadget quality is not expressed by those metrics. Due to this lack of related work, we introduce the metrics that are currently used to evaluate CFI implementations, and discuss gadgets required for carrying out attacks against CFI.

Zhang et al.~\cite{cfi_for_cots} propose using AIR which denotes how many gadgets are removed, because they are not acceptable targets of indirect branches. However, AIR does not take into account the quality of the remaining gadgets. Payer et al.~\cite{lockdown} propose DAIR, which works similarly to AIR but is dynamic, hence varies during program execution. Tice at al.~\cite{forward_edge_cfi} propose forward-edge AIR (fAIR), which is computed like AIR, but takes into account only forward-edge indirect control transfers, i.e., calls and jumps. All of the above metrics are limited to CFI though, and do not consider the quality of the remaining gadgets.

Carlini et al.~\cite{cfi_bending} discuss the effectiveness of CFI implementations against ROP attacks and propose what they call a basic exploitation test (BET). BET consists of three generalized attack scenarios, namely arbitrary code execution, confined code execution and information leakage. They use a minimal program that allows exploitation, apply several CFI implementations, and evaluate, which of the described attack scenarios could be achieved. However, this process was done by a human, hence dependant on skill and knowledge of the exploit developer. Therefore, it can also not be used for mass-analyzing binaries.

In 2014 and 2015 many attacks targeting various CFI implementations, e.g., kBouncer~\cite{kbouncer}, ROPecker~\cite{ropecker}, or CFI for COTS~\cite{cfi_for_cots} have been published. As CFI places tight restrictions on indirect control-flow transfers, hence also gadgets, those attacks often incorporate gadgets that would rarely be used in real attacks. E.g., Carlini and Wagner~\cite{rop_dangerous}, Davi et al.~\cite{stitching_gadgets}, and G\"{o}kta\c{s} et al.~\cite{size_matters} discovered that long gadgets with few side effects are suitable for breaking heuristics-based mitigations. Such gadgets should consist of at least 20 instructions, preserve as many registers as possible, have few side-effects, and easily fulfillable preconditions. Gadgets of this length are generally not useful in today's attacks, and therefore GaLity does not treat them any different than other gadgets.
Another kind of gadget commonly used in these attacks is an LBR-flushing gadget~\cite{rop_dangerous, antirop_evaluation}. Recent CPUs have special registers which can be configured to store the addresses of up to the 16 most recent taken indirect branches~\cite{intel_architecture_manual}, which is a feature kBouncer~\cite{kbouncer} and ROPecker~\cite{ropecker} use. When certain, critical APIs are invoked, the LBR is inspected and, depending on whether the control-flow appears legitimate or not, an exception is raised. LBR-flushing gadgets are gadgets that naturally contain many indirect branches, present in the regular control flow, e.g., functions that call lots of sub-functions. By using such a gadget, the LBR is filled with legitimate addresses and there is no trace of irregular control flow, i.e., ROP, in the LBR.

Q~\cite{q} allows exploit developers to write a target program in the high-level language QooL and automatically builds a ROP chain that uses only gadgets from a provided binary. However, Q handles gadgets with side effects, which we call preconditions in this paper, very conservatively and discards such gadgets, potentially removing a large number of useful gadgets. Homescu et al.~\cite{microgadgets} present a Turing-complete set of gadgets, using only gadgets that are 3 byte or shorter. They find that all required gadgets appear very frequently in regular Linux binaries.

Lastly, there are many tools that assist exploit developers by finding and sorting gadgets, but none of them take into account the quality of gadgets. Some of these tools also attempt to automatically build a ROP exploit for one predefined scenario {e.g., ROPgadget\footnote{\url{http://shell-storm.org/project/ROPgadget/}}, Mona.py\footnote{\url{https://www.corelan.be/index.php/2011/07/14/mona-py-the-manual/}}, or ropper\footnote{\url{https://scoding.de/ropper/}}, however, from our experience they are not very sophisticated and often fail, even if the necessary gadgets are available.

\section{Conclusion}
\label{Sec:Conclusion}
Return-Oriented Programming forms the cornerstone of many contemporary exploitation techniques, yet its viability hinges on the availability of useful gadgets. Program transformations, including exploit mitigation techniques, often do not take into consideration their impact on the quality and number of gadgets that they introduce into the binary to which they are applied. Evaluations usually concentrate on the security gained, but not the security that might be lost due to a set of gadgets that is now favourable for an attacker than in the original, unmodified binary.
\newline
This work addresses this issue and allows researchers to consider this important aspect, by developing a set of metrics that, by combining concrete attack scenarios and measuring overall gadget quality, cover a wide range of possible exploit scenarios. We implemented the described metrics in a tool called \tool{}, and applied it to binaries compiled with MPX, a new buffer overflow prevention technique introduced by Intel. Our results show that MPX provides gadgets of higher quality, and also a favourable set of gadgets in a concrete attack scenario.

\section*{Acknowledgements}
We want to express our thanks to the anonymous reviewers for their valuable comments. In particular, we want to thank our shepherd, Mathias Payer, who helped us give this paper its final form. This work was supported by the BMBF within EC SPRIDE, by the Hessian LOEWE excellence initiative within CASED, by the DFG Collaborative Research Center CROSSING, by the DFG Priority Program 1496 Reliably Secure Software Systems, and the project INTERFLOW.

\bibliographystyle{abbrv}
\bibliography{papers}

\clearpage

\end{document}